\documentclass[letterpaper]{article} 
\usepackage{aaai25}  
\usepackage{times}  
\usepackage{helvet}  
\usepackage{courier}  
\usepackage[hyphens]{url}  
\usepackage{graphicx} 
\usepackage{natbib}  
\usepackage{caption} 
\usepackage{algorithm}
\usepackage{graphicx}
\usepackage{subcaption}
\usepackage{colortbl}
\usepackage{booktabs}
\usepackage{multirow}
\usepackage[T1]{fontenc}
\usepackage{amsmath}    
\usepackage{multirow}

\usepackage{amsfonts}
\usepackage{stackrel}   
\usepackage{float}
\usepackage{booktabs}
\usepackage{multirow}%
\usepackage{amsmath}%
\usepackage{amsthm}%
\usepackage{mathrsfs}%
\usepackage[title]{appendix}%
\usepackage{xcolor}%
\usepackage{textcomp}%
\usepackage{manyfoot}%
\usepackage{booktabs}%
\usepackage{algpseudocode}%
\usepackage{listings}%
\usepackage{amsmath}    
\usepackage{amsfonts}
\usepackage{stackrel}  
\usepackage{adjustbox}
\usepackage{xcolor} 
\usepackage{amsmath, amsfonts, amsthm, stackrel}   
\usepackage{ragged2e} 
\usepackage{booktabs,makecell, multirow, tabularx}

\usepackage[figuresright]{rotating}
\usepackage{newfloat}
\usepackage{listings}
\DeclareCaptionStyle{ruled}{labelfont=normalfont,labelsep=colon,strut=off} 
\floatstyle{ruled}
\newfloat{listing}{tb}{lst}{}
\floatname{listing}{Listing}
\title{Enhance Large Language Models as Recommendation Systems \\ with Collaborative Filtering}
\author{
    Zhisheng Yang\textsuperscript{\rm 1},
    Xiaofei Xu\textsuperscript{\rm 2},
    Ke Deng\textsuperscript{\rm 3}\thanks{Corresponding author},
    Li Li\textsuperscript{\rm 1}\thanks{Corresponding author}
}
\affiliations{
    \textsuperscript{\rm 1}College of Computer and Information Science, Southwest University, Chongqing, 400715, China\\
    \textsuperscript{\rm 2}School of Information Technology, Murdoch University, Murdoch, WA 6150, Australia\\
    \textsuperscript{\rm 3}School of Computing Technologies, RMIT University, Melbourne, VIC 3000, Australia \\
    yzsheng@email.swu.edu.cn,\;
    xiaofei.xu@murdoch.edu.au,\;
    ke.deng@rmit.edu.au,\;
    lily@swu.edu.cn
}

\usepackage{bibentry}

\begin{document}

\maketitle

\begin{abstract}
As powerful tools in Natural Language Processing (NLP), Large Language Models (LLMs) have been leveraged for crafting recommendations to achieve precise alignment with user preferences and elevate the quality of the recommendations. The existing approaches implement both non-tuning and tuning strategies. Compared to following the tuning strategy, the approaches following the non-tuning strategy avoid the relatively costly, time-consuming, and expertise-requiring process of further training pre-trained LLMs on task-specific datasets, but they suffer the issue of not having the task-specific business or local enterprise knowledge. To the best of our knowledge, none of the existing approaches following the non-tuning strategy explicitly integrates collaborative filtering, one of the most successful recommendation techniques. This study aims to fill the gap by proposing critique-based LLMs as recommendation systems (Critic-LLM-RS). For our purpose, we train a separate machine-learning model called \textit{Critic} that implements collaborative filtering for recommendations by learning from the interactions between many users and items. The Critic provides critiques to LLMs to significantly refine the recommendations. Extensive experiments have verified the effectiveness of Critic-LLM-RS on real datasets.
\end{abstract}

%
\section{Introduction}
As powerful tools in Natural Language Processing (NLP), Large Language Models (LLMs) are causing significant impact in the field of recommendations (see a survey \cite{wu2023survey}). 
In this field, LLMs as recommendation systems (\textit{LLM-as-RS}) refers to the approaches that directly leverage LLMs for crafting recommendations to achieve precise alignment with user preferences and elevate the quality of the recommendations. 

\textit{LLM-as-RS} has been implemented in tuning or non-tuning strategies. Following the tuning strategy, models meticulously calibrate the alignment of LLMs with specific recommendation tasks through targeted training or fine-tuning via methods like instruction tuning, content-based fine-tuning, or prompt adjustment. This strategy is exemplified by models such as TALLRec \cite{bao2023tallrec}, FLAN-T5 \cite{kang2023llms}, LLMs fine-tuned expressly for content recommendations \cite{xu2024enhancing}, RecSysLLM \cite{chu2023leveraging}, and POD \cite{li2023prompt,liu2023pre}. Following the non-tuning strategy, models rely on the inherent, pre-trained prowess of LLMs to directly craft recommendations via meticulously designed prompts. This method bypasses the need for any model adjustments or training, rendering it perfectly suited for scenarios like zero-shot or few-shot learning. This strategy isn't limited to models like ChatGPT \cite{kocon2023chatgpt}, and LLaMA but also encompasses 
Llama4Rec \cite{luo2024integrating}, InteraRec \cite{karra2024interarec}, GENREC \cite{ji2024genrec}, FaiRLLM \cite{zhang2023chatgpt}, and GPTRec \cite{liu2023chatgpt,petrov2023generative}.

Compared to those following the tuning strategy, the approaches following the non-tuning strategy avoid the relatively costly, time-consuming, and expertise-requiring process of further training pre-trained LLMs on task-specific datasets, but they suffer the issue of not having the required task-specific business or local enterprise knowledge. To the best of our knowledge, none of the existing approaches following the non-tuning strategy explicitly integrate collaborative filtering, one of the most successful recommendation technologies\cite{8506344}. Collaborative filtering separates out items that a user might like based on reactions by similar users. It works by searching for a large group of people and finding a smaller set of users with tastes similar to a particular user. 

To fill the research gap, this study proposes \textit{critique-based LLMs as recommendation systems} (Critic-LLM-RS). It is motivated by a recent method developed in natural language processing, i.e., providing textual feedback/critiques on the generated response of LLM and then prompting the LLM to refine the response \cite{schick2022peer,akyurek2023rl4f,lee2023rlaif,yu2023improving}. In these studies, a separate, usually small, LLM model is trained to learn how to generate feedback/critiques as a human or domain expert. For our purpose, we train a separate machine-learning model called \textit{Recommendation Critic} (R-critic) that implements collaborative filtering for recommendations by learning from interactions between many users and items. The R-critic provides critiques on the initial recommendations by LLMs so that LLMs can refine the final recommendation. Critic-LLM-RS circumvents further training pre-trained LLMs but possesses task-specific knowledge, i.e., the capability of collaborative filtering. The contributions of this study are threefold.
\begin{itemize}
    \item This study is the first, to the best of our knowledge, that explicitly integrates collaborative filtering, one of the most important recommendation technologies, into \textit{LLM-as-RS} where the non-turning strategy is followed.     
    \item This study implements collaborative filtering by a separate model R-critic to provide critiques on the initial recommendations of LLMs and prompt LLMs to refine the final recommendations.  
    \item This study has verified the effectiveness of the proposed Critic-LLM-RS by case studies and extensive tests against the existing non-tuning LLM-as-RS approaches on real datasets.  
\end{itemize}

\begin{figure*}[h]
  \centering
  \includegraphics[width=0.8\linewidth]{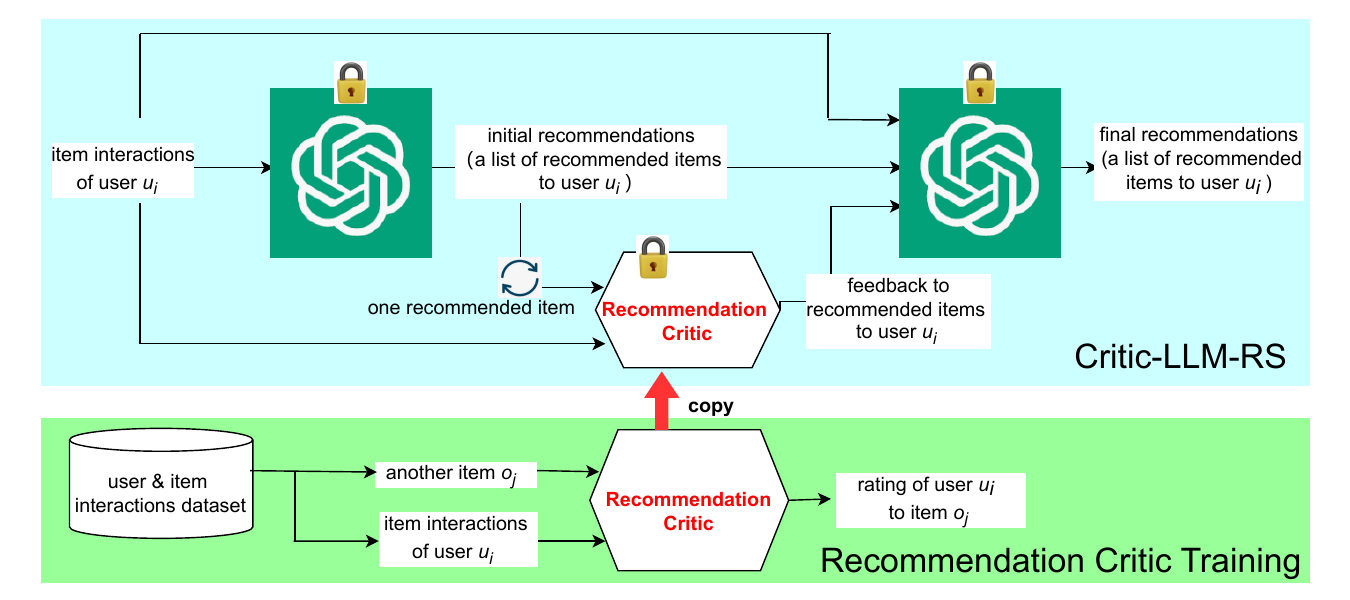}
  \caption{Critic-LLM-RS Architecture}
  \label{fig:system}
\end{figure*}
\section{Related work}
\subsection*{LLMs for Recommendations}\label{sec:relatedLLMs}




Large Language Models (LLMs) are causing a significant impact in the research field of recommendations. One line of research is Discriminative LLMs for Recommendation (DLLM4Rec)\cite{wu2023survey}. These approaches typically leverage BERT and its variants for extracting and embedding high-quality textual features and comprehensive knowledge, which are then integrated with recommender systems to enhance recommendations\cite{devlin2018bert,wu2024exploring,qiu2021u,yang2022lightweight,wu2022multi,hou2022towards,wu2021empowering,yang2022lightweight,liu2022boosting,li2023exploring,yuan2023go,zhang2024agentcf}. 

Another line of research is Generative LLMs for Recommendation (GLLM4Rec)\cite{wu2023survey}. These approaches utilize the language generation capabilities of LLMs to directly handle the recommendation tasks. GLLM4Rec operates in two main modes: \textit{LLM Tokens+RS} and \textit{LLM as RS}\cite{lin2023can,trichopoulos2023large}. The \textit{LLM Tokens+RS} approach leverages semantic mining techniques to uncover potential user preferences by synthesizing tokens generated from item and user characteristics, and then integrating these insights into the recommendations\cite{wei2024llmrec,zheng2023generative,wu2024exploring,xi2023towards,wang2023generative,wang2022towards,xu2024enhancing,wang2023recagent}. The \textit{LLM as RS} approach directly leverages LLMs for crafting recommendations to achieve precise alignment with user preferences and elevate the quality of the recommendations\cite{wu2023survey}.   

The approaches of \textit{LLM as RS} implement both non-tuning and tuning strategies. Following the tuning strategies, models meticulously calibrate the alignment of LLMs with specific recommendation tasks through targeted training or fine-tuning via methods like instruction tuning, content-based fine-tuning, or prompt adjustment. This strategy is exemplified by models such as TALLRec \cite{bao2023tallrec}, FLAN-T5 \cite{kang2023llms}, LLMs fine-tuned expressly for content recommendations \cite{xu2024enhancing}, RecSysLLM \cite{chu2023leveraging}, and POD \cite{li2023prompt,liu2023pre}. However, further training pre-trained LLMs on task-specific datasets is a relatively costly, time-consuming, and expertise-requiring process. Following the non-tuning strategy, models rely on the inherent, pre-trained prowess of LLMs to directly craft recommendations via meticulously designed prompts \cite{luo2024integrating, karra2024interarec, ji2024genrec, zhang2023chatgpt, liu2023chatgpt,petrov2023generative}. This method bypasses the need for any model adjustments or training, rendering it perfectly suited for scenarios like zero-shot or few-shot learning \cite{hou2024large,che2024new,10.1145/3604915.3608845}. In \cite{wang2023recmind}, it breaks complex recommendation tasks into manageable steps, each step involves thought, action, and observation, and the agent considers all previously explored paths for the next planning.

\subsection*{LLMs with Feedback}
Recently, some researchers focused on providing textual feedback on the generated response and letting LLM refine the response~\cite{schick2022peer,akyurek2023rl4f,lee2023rlaif,yu2023improving}. Such approaches allow for the refinement of model response, improving the quality of response and reducing the likelihood of hallucinations. An early research of critic-style feedback is directly asking humans for feedback. Since the feedback is provided by humans, it would be expensive to collect the feedback, so the generalization ability is very limited. Another way is to utilize a critique model to generate those critiques~\cite{schick2022peer,akyurek2023rl4f}, they will collect gold standard critiques from online forums or Wikipedia edit histories as training data and then use supervised fine-tuning or reinforcement learning to obtain a critique model, typically a small LLM. However, when applied to a specific domain, such datasets may simply not exist or may be very costly to acquire. In \cite{kalyanpur2024llm}, an Automated Reasoning engine is used as the Critic to enhance the logical reasoning capabilities of LLMs.


\subsection*{Remarks}
Many datasets have been collected for various recommendations tasks\footnote{https://github.com/caserec/Datasets-for-Recommender-Systems}. By exploring the relevant dataset, this study effectively harnesses the capability of collaborative filtering by a separate machine-learning model, known as \textit{Recommendation Critic}. 
It provides implicit user feedback to boost LLMs as a recommendation system following the light-weight non-tuning strategy. 

\section{Methodology}

\subsection{Critic-LLM-RS Architecture}
The architecture of the proposed \textit{critique-based LLMs as a recommendation system} (Critic-LLM-RS) is presented in Figure \ref{fig:system}. Critic-LLM-RS consists of a pre-trained LLM and a pre-trained machine learning model \textit{Critic} where the lock sign indicates the model parameters are frozen. The input of LLM entails the interaction history of user $u_i$ with items (such as a list of movies that the user watched and rated previously) and a prompt to indicate the task is to recommend items that $u_i$ likes. The interaction with an item includes the item name (e.g., movie name), item information if available (e.g., director and actors), and the rating given by $u_i$ (e.g., 1-5). The LLM outputs the initial recommendations, denoted as $\Lambda$. 

Then each recommended item is evaluated by the Recommendation Critic. The input of the Recommendation Critic includes a recommended item $o\in \Lambda$ and the interaction history of user $u_i$ with items. The Recommendation Critic has been pre-trained for the capability of collaborating filtering. With this capability, the Recommendation Critic provides feedback to the recommended item $o$ based on the ratings of other users with similar preferences for the same or similar item(s). The feedback to all recommended items in $\Lambda$ is then returned to the LLM for an improved recommendation for users $u_i$.    

\subsection{Recommendation Critic}
The Recommendation Critic is a supervised model pre-trained to evaluate each of the items in initial recommendations by LLM. Training data consists of users, each with an interaction history with items. Each interaction entails the item name (e.g., movie name), item information if available (e.g., director and actors), and the rating given by the user (e.g., 1-5). Based on the interaction history, the Recommendation Critic is trained to estimate the user's rating to an item $o$ against the real rating. Specifically, the textual information in interaction history is pre-processed and embedded using the pre-trained BERT\footnote{https://huggingface.co/google-bert/bert-base-uncased}. The textual information of the item $o$ is embedded in the same way. The embeddings of history interactions and the embeddings of $o$ are input to the Recommendation Critic which outputs the estimated rating. The parameters of the Recommendation Critic are optimized by minimizing the difference between the estimated and real ratings. Without loss of generality, the Recommendation Critic is implemented as a multiclass classifier where each class corresponds to a rating level.      

As one of the most successful technologies for building recommender systems, collaborative filtering uses the known preferences of a group of users to make recommendations or predictions of the unknown preferences for other users \cite{He05031}. The underlying assumption is that if a user $A$ often has the same ratings as a user $B$ on items previously, $A$ is likely to have $B$'s rating on a different item than that of a randomly chosen person. The Recommendation Critic model has the capability of collaborative filtering. If two users have similar preferences represented by interaction history, the Recommendation Critic model is trained so that the estimated ratings they give to the same item are similar.    

Moreover, the Recommendation Critic is relevant to content-based filtering when the items are not simply represented by item identities but by item features (i.e., textual information). Content-based filtering uses item features to recommend other items similar to what the user likes, based on their previous actions or explicit feedback \cite{Meteren2000UsingCF}. For simplicity of presentation, we state the Recommendation Critic possesses the capability of collaborative filtering only in this study.        



\begin{table*}[t!]
\footnotesize
\centering
\caption{A case study - Movies dataset - real (real rating), Critic (rating estimated by the Recommendation Critic)}
\label{tbl:case1}
\begin{tabular}{p{6.9cm}|p{0.35cm}|p{0.6cm}||p{6.5cm}|p{0.35cm}|p{0.6cm}}
\toprule\toprule
\multicolumn{6}{p{17.5cm}}{\includegraphics[width=\textwidth]{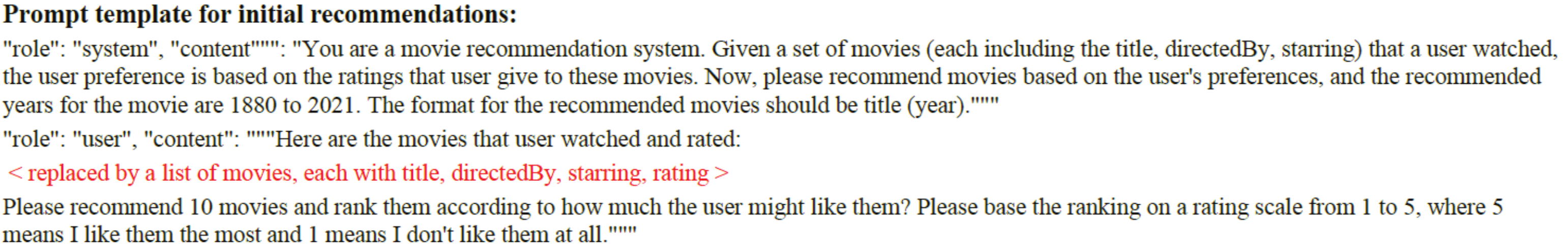}}
\\ \bottomrule \bottomrule
\multicolumn{6}{p{17.5cm}}{\includegraphics[width=\textwidth]{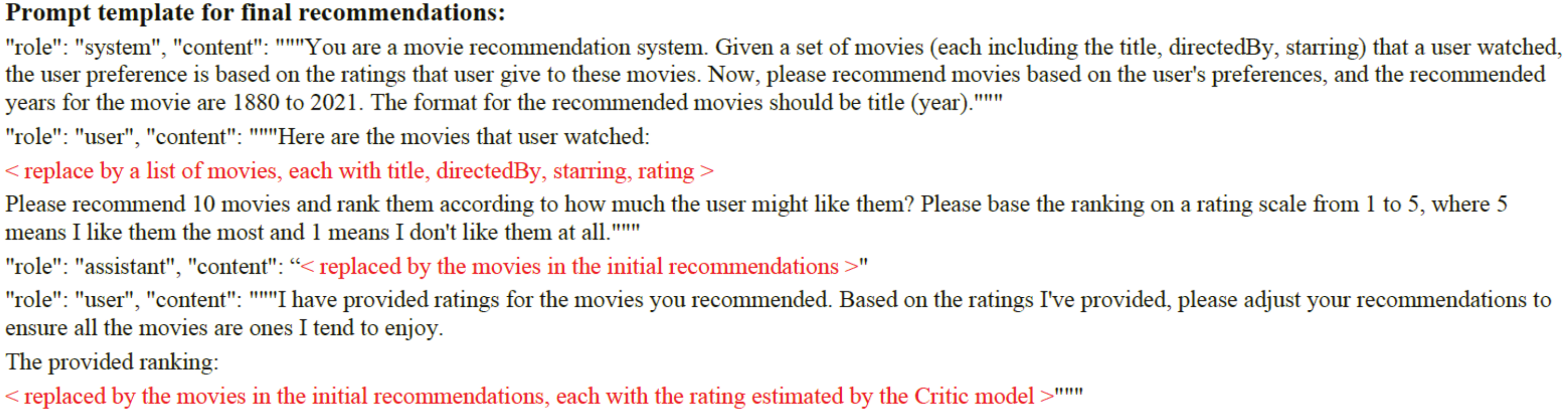}}
\\ \bottomrule \bottomrule

\multicolumn{6}{p{17.5cm}}{\textbf{Movies watched and rated by the user:} {[} "\textbf{title}": "Airheads (1994)", "\textbf{directedBy}": "Michael Lehmann",  "\textbf{starring}": "Steve Buscemi, Chris Farley, Brendan Fraser, Adam Sandler, Joe Mantegna, David Arquette, Amy Locane, Ernie Hudson, Judd Nelson, Michael McKean, Michael Richards", "\textbf{rating}": "4.0"\},\{"\textbf{title}": "Red Dragon (2002)", "\textbf{directedBy}": "Brett Ratner", "\textbf{starring}": "Ralph Fiennes, Anthony Hopkins, Harvey Keitel, Edward Norton, Philip Seymour Hoffman, Mary-Louise Parker, Ken Leung", "\textbf{rating}": "3.5"\}, $\dots${]}}
\\ \bottomrule 
\textbf{Initial recommendations} & \textbf{real} & \textbf{Critic} & \textbf{Final Recommendations} & \textbf{real} & \textbf{Critic}
\\ \bottomrule 
\begin{tabular}[c]{@{}l@{}}\textcolor{blue}{1 Jurassic Park (1993)}\\ 2 \textcolor{blue}{Gladiator (2000)}\\ 3 The Dark Knight (2008)\\ 4 Fast and Furious 7 (2015)\\ \textcolor{blue}{5 Die Hard (1988)}\\ 6 The Lord of the Rings: The Return of the King (2003)\\ \textcolor{blue}{7 Blade (1998)}\\ \textcolor{blue}{8 Pulp Fiction (1994)}\\ 9 The Shawshank Redemption (1994)\\ 10 John Wick (2014)\end{tabular}  
& \begin{tabular}[c]{@{}l@{}}4.0 \\ 3.0 \\- \\- \\5.0 \\- \\2.0 \\4.0 \\- \\ - \end{tabular}&\begin{tabular}[c]{@{}l@{}}4.0 \\ 2.5 \\3.0 \\4.5 \\5.0 \\3.5 \\2.0 \\4.0 \\3.5 \\ 3.5\end{tabular}  &
\begin{tabular}[c]{@{}l@{}}\textcolor{blue}{1 Pulp Fiction (1994)}\\ 2 Predator (1987)\\ \textcolor{blue}{3 Terminator 2: Judgment Day (1991)}\\ \textcolor{blue}{4 Dr. Strangelove or: How I Learned to Stop (1964)}\\ 5 The Night of the Living Dead (1968)\\ \textcolor{blue}{6 Apocalypse Now (1979)} \\ 7 The Terminator (1984): The Winter Soldier (2014)\\ 8 Spider-Man: Into the Spider-Verse (2018)\\ 9 Gravity (2013)\\ 10 Twin Peaks: Fire Walker Stunt Double (2015)\end{tabular}
& \begin{tabular}[c]{@{}l@{}} 4.0 \\ -\\ 4.5\\5.0\\- \\ 5.0 \\-\\ -\\ - \\ -\end{tabular} & 
\begin{tabular}[c]{@{}l@{}}  4.0\\ 3.0\\ 5.0\\ 5.0\\  3.5\\  5.0 \\  4.0\\  3.0\\  3.5\\ 3.0\end{tabular} \\ 
\bottomrule \bottomrule 
\end{tabular}
\end{table*}

\section{Motivation of Critic-LLM-RS}
Given the training dataset, one may ask two questions: (1) why not train a recommendation system using one of the traditional recommendation technologies? and (2) why not directly fine-tune the LLM model?  Regarding question one, the traditional recommendation systems typically depend on the training dataset. The system can only recommend the items in the training dataset. However, it is hard to maintain a dataset containing all items in a particular recommendation task such as movies all the time. Regarding question two, LLM fine-tuning is a supervised learning process where one uses a dataset of labeled examples to update the parameters of LLM and make the model improve its ability for specific tasks. It is a powerful technique, but it comes with some notable drawbacks \cite{wu2023survey}. First, LLMs have millions or even billions of parameters. Training these massive models requires serious computational power. For smaller teams or those on a budget, the costs can quickly become prohibitive. Second, fine-tuned models can struggle to adapt to new information without expensive retraining. Third, fine-tuning LLMs needs specialized skills and knowledge that can be hard to find. Finally, fine-tuning LLMs models can sometimes "hallucinate" strange or biased results, or completely forget their previous training. 

By combining the pre-trained LLM with the pre-trained Recommendation Critic, the Critic-LLM-RS enjoys the advantages of LLM and traditional recommendation systems. First, LLMs store an extensive amount of factual knowledge obtained from vast collections of documents. It knows large and latest datasets in a wide range of fields so that LLMs can recommend the items not in the training dataset. 
Second, the Recommendation Critic harnesses collaborative filtering, one of the most important technologies in traditional recommendation systems. Following a way that has been verified effective in improving LLM responses for particular tasks, the Critic-LLM-RS inserts the feedback from the Recommendation Critic into LLM for a better recommendation. Because pre-trained and updated separately, various Recommendation Critic models can be built up on datasets for different recommendation tasks, and work with the Critic-LLM-RS in the plug-in and plug-out mode.  


\section{A Case Study} 
In Table \ref{tbl:case1},
the recommendations of two users are reported for Movies, respectively (see details of datasets in the Experiment section). The prompt provided to LLM for the initial recommendations and the prompt for the final recommendations are also illustrated. For the user in Table \ref{tbl:case1}, the recommendations in black font are the movies not having real ratings from the user, and in blue font are the movies having real ratings. Compared with the initial recommendations, we can observe that (1) the real ratings of movies in the final recommendations increase, and (2) the ranking order of the movies is improved, i.e., the movies with the higher real and estimated ratings are closer to the top of the rank. 

\begin{figure}[t]
\centering
\begin{subfigure}[b]{\columnwidth}
    \centering
    \includegraphics[width=0.8\columnwidth]{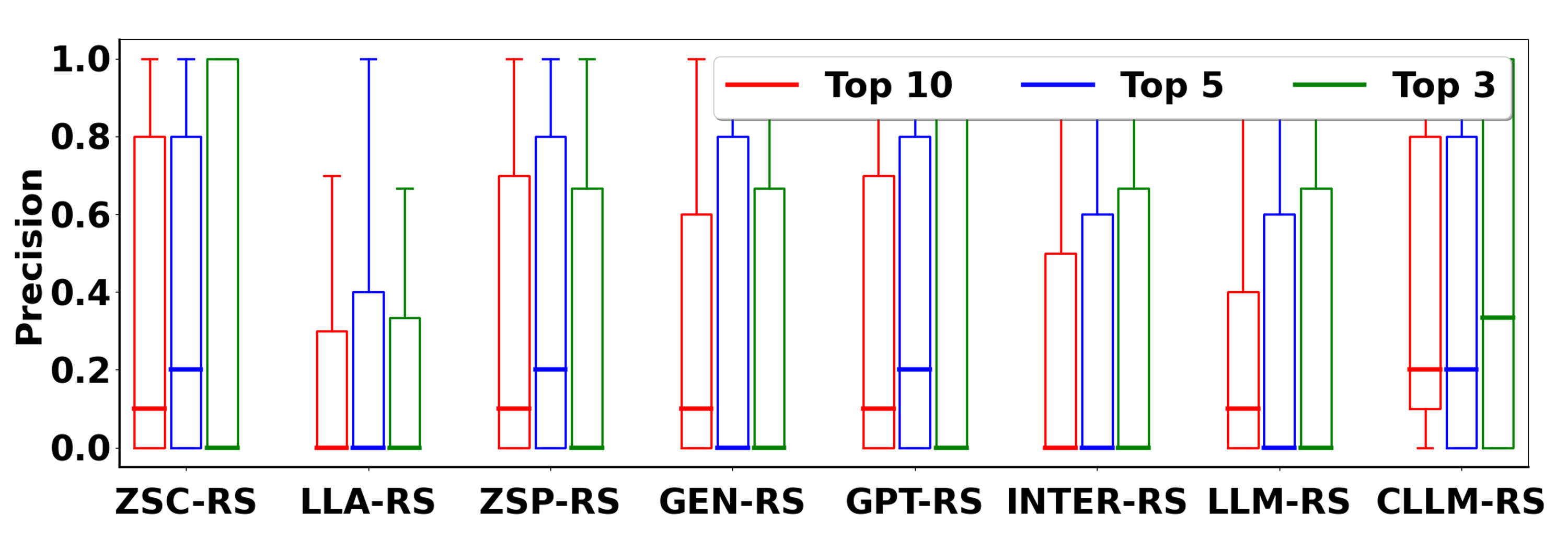}
    \label{fig:precision_book_baselines}
\end{subfigure}

\begin{subfigure}[b]{\columnwidth}
    \centering
    \includegraphics[width=0.8\columnwidth]{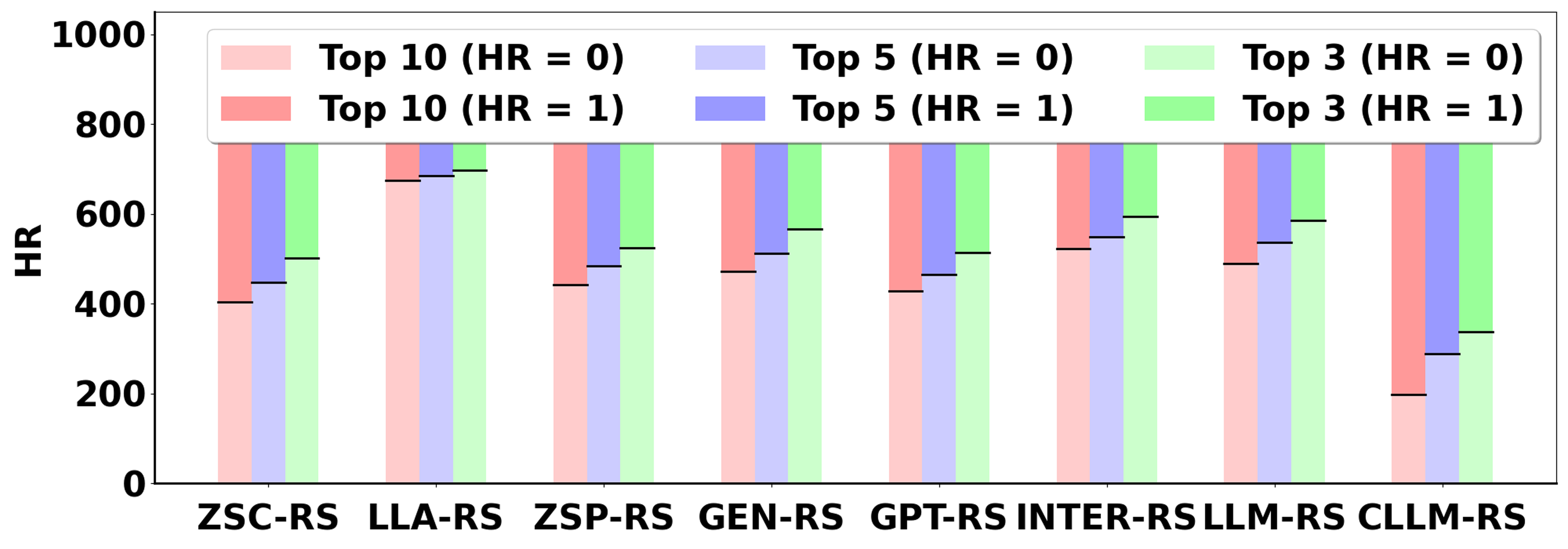}
    \label{fig:hr_book_baselines}
\end{subfigure}

\begin{subfigure}[b]{\columnwidth}
    \centering
    \includegraphics[width=0.8\columnwidth]{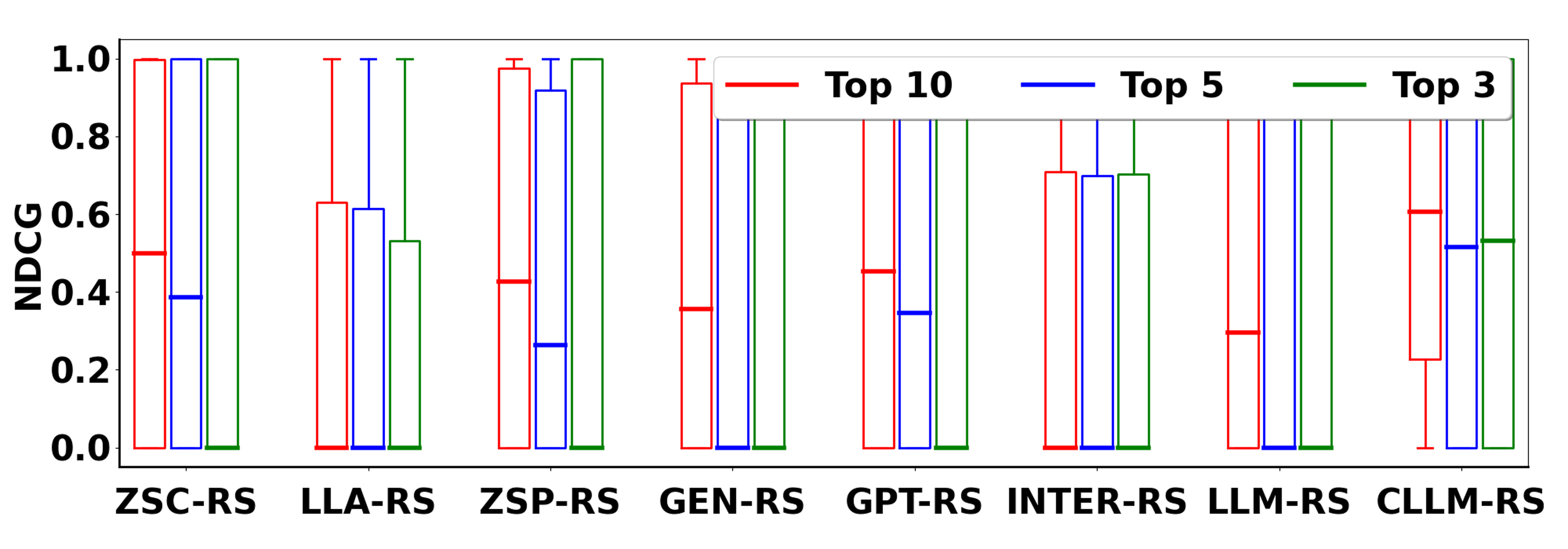}
    \label{fig:ndcg_book_baselines}
\end{subfigure}
\caption{Evaluation on Books with Oracle.}
\label{fig:book_baselines}
\end{figure}



\begin{figure}[t]
\centering
\begin{subfigure}[b]{\columnwidth}
    \centering
    \includegraphics[width=0.8\columnwidth]{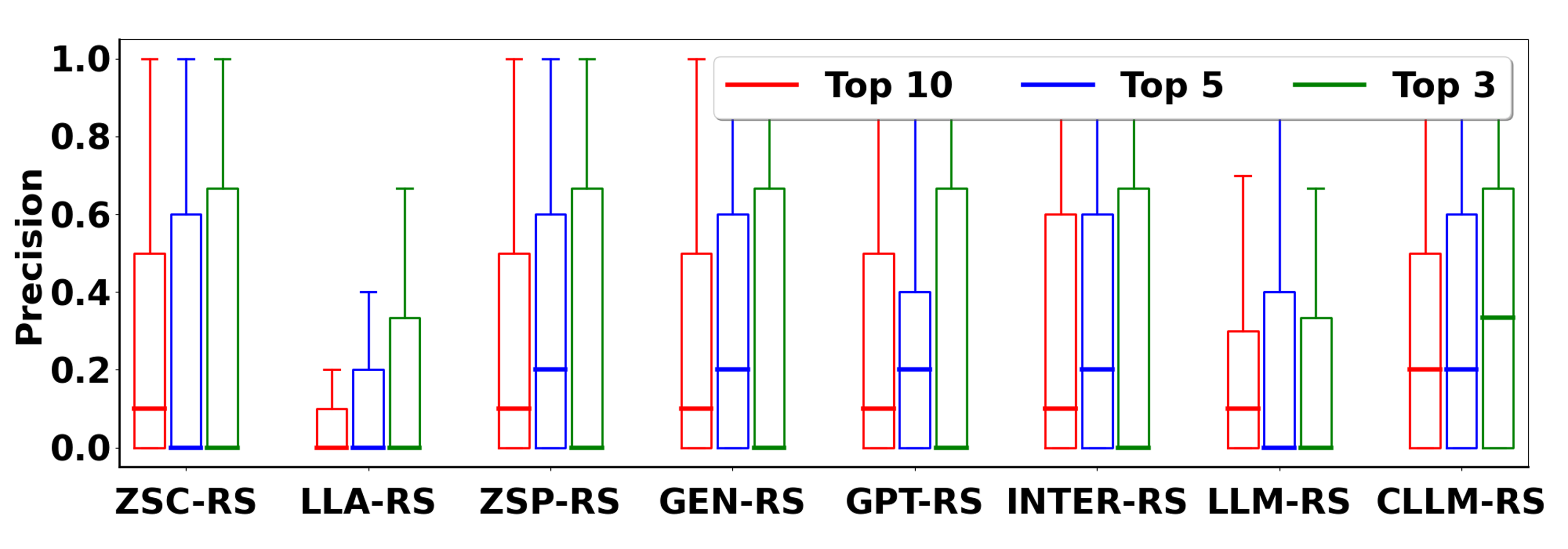}
    \label{fig:precision_movie_baselines}
\end{subfigure}

\begin{subfigure}[b]{\columnwidth}
    \centering
    \includegraphics[width=0.8\columnwidth]{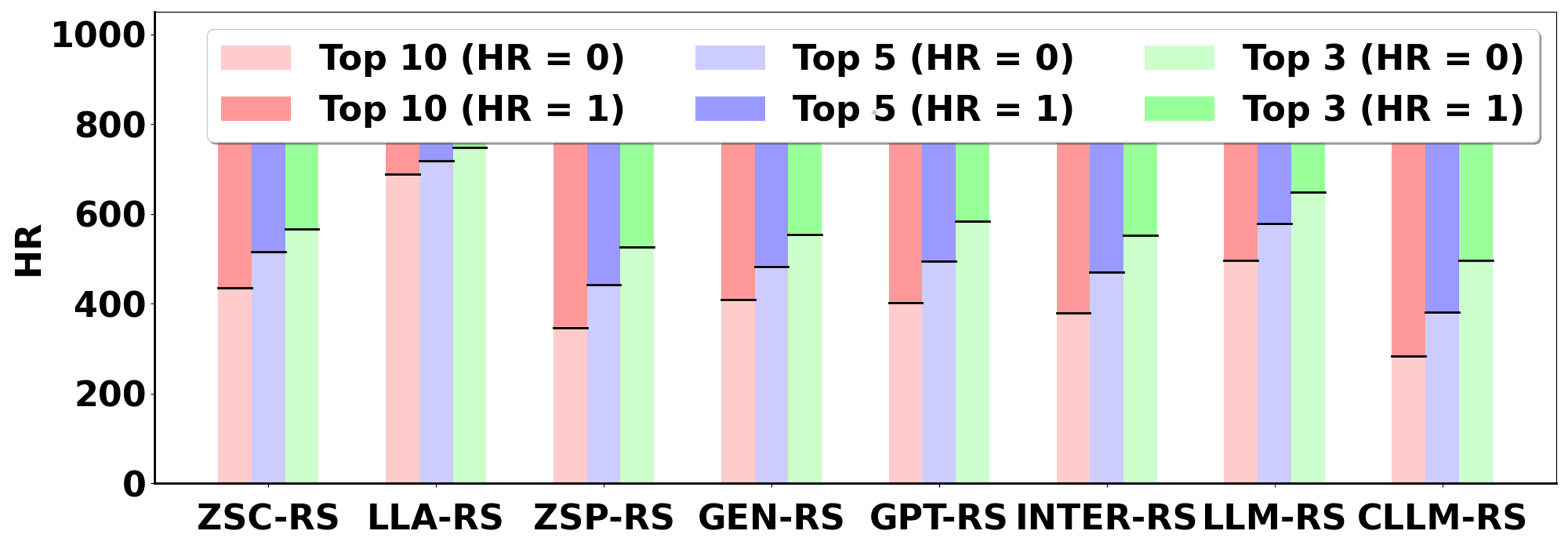}
    \label{fig:hr_movie_baselines}
\end{subfigure}

\begin{subfigure}[b]{\columnwidth}
    \centering
    \includegraphics[width=0.8\columnwidth]{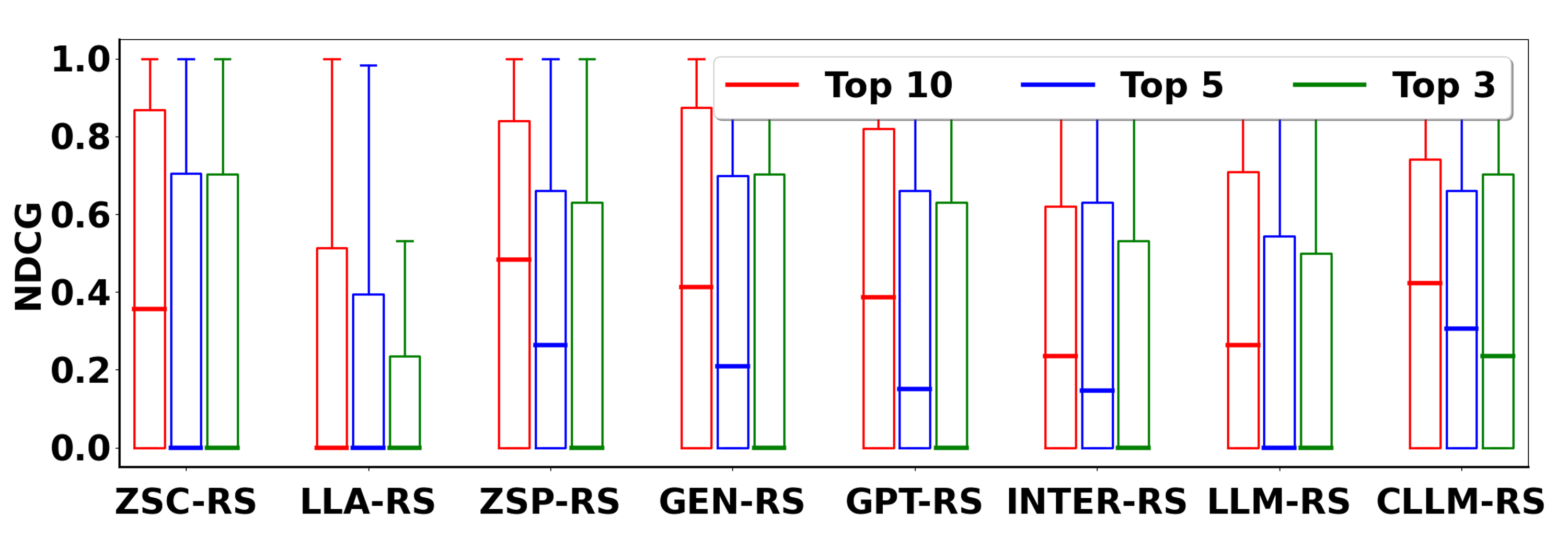}
    \label{fig:ndcg_movie_baselines}
\end{subfigure}
\caption{Evaluation on Movies with Oracle.}
\label{fig:movie_baselines}
\end{figure}


\begin{figure}[t]
\centering
\begin{subfigure}[b]{\columnwidth}
    \centering
    \includegraphics[width=0.8\columnwidth]{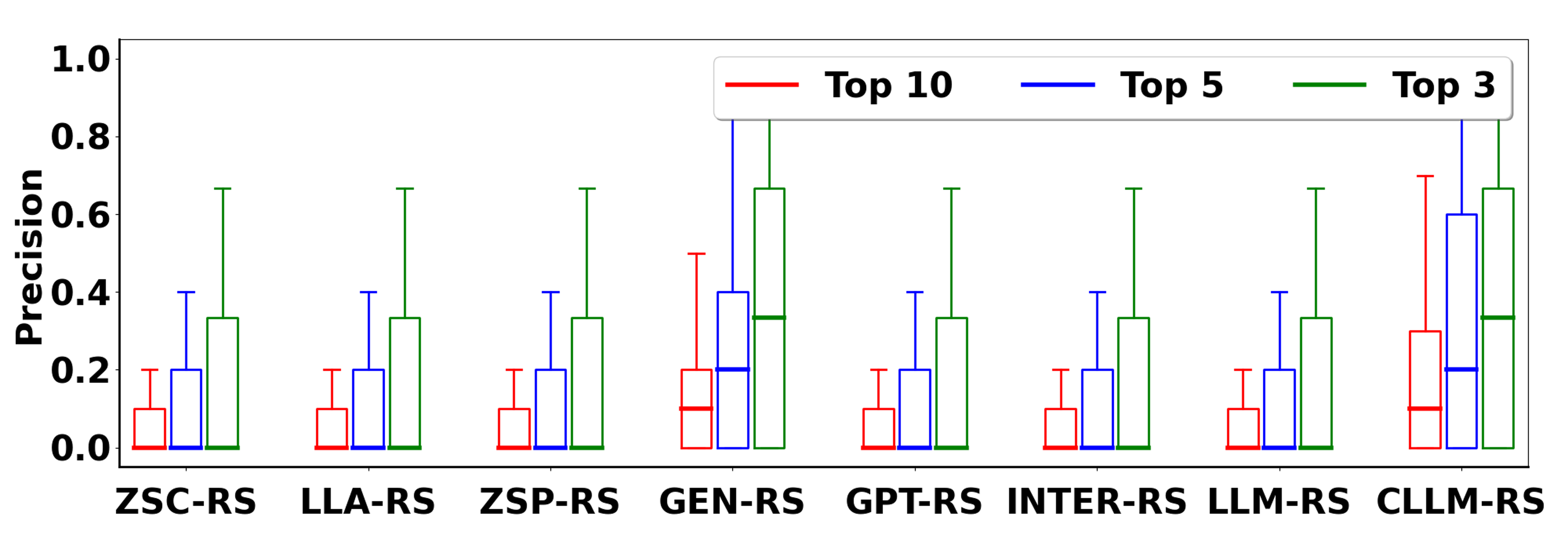}
    \label{fig:precision_book_candidate}
\end{subfigure}

\begin{subfigure}[b]{\columnwidth}
    \centering
    \includegraphics[width=0.8\columnwidth]{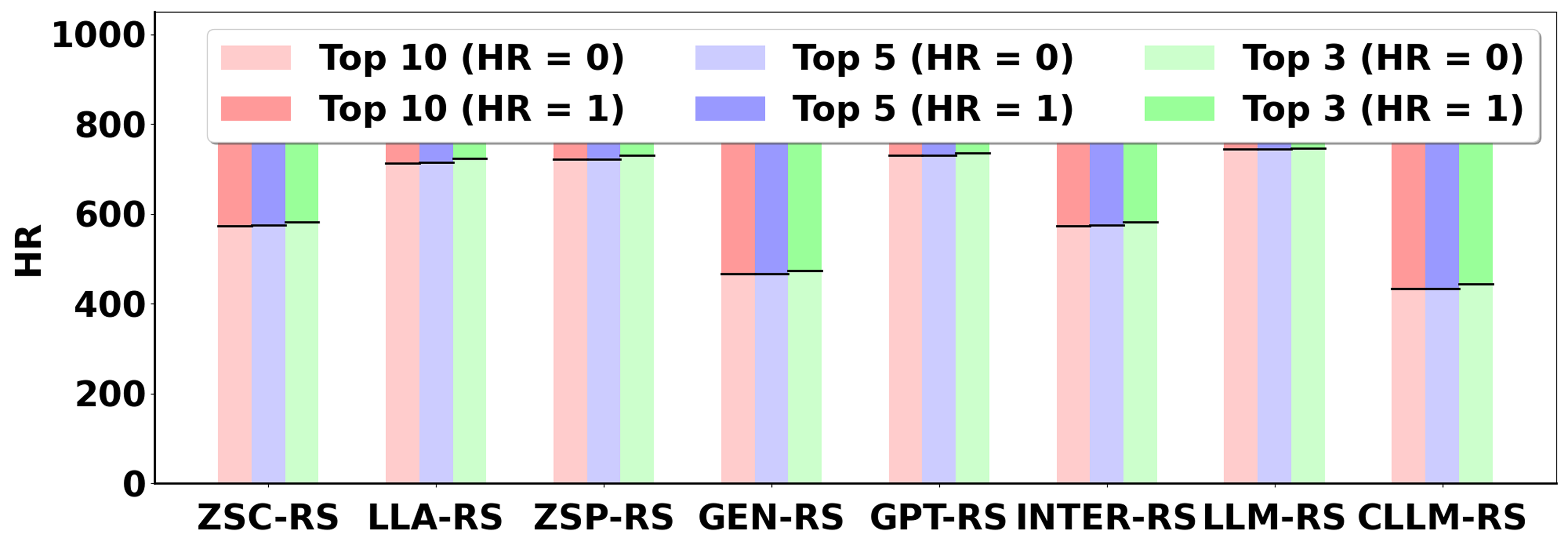}
    \label{fig:hr_book_candidate}
\end{subfigure}

\begin{subfigure}[b]{\columnwidth}
    \centering
    \includegraphics[width=0.8\columnwidth]{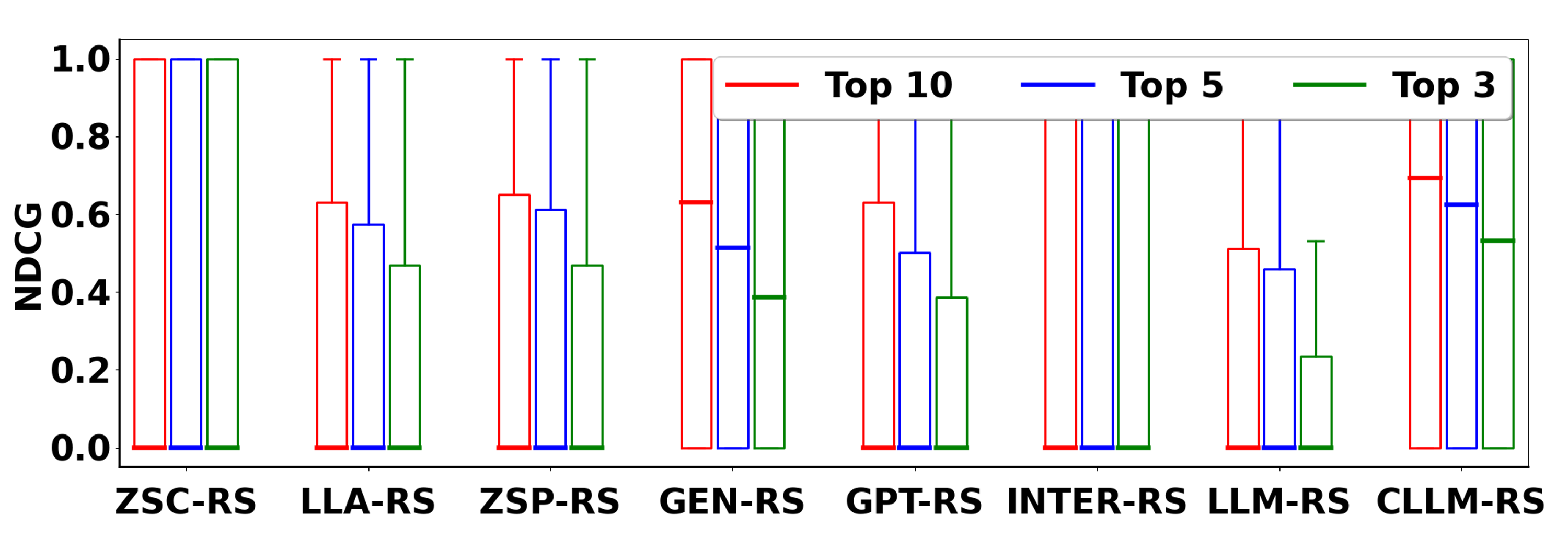}
    \label{fig:ndcg_book_candidate}
\end{subfigure}
\caption{Evaluation on Books with a candidate set.}
\label{fig:book_critic_candidate}
\end{figure}

\begin{figure}[t]
\centering
\begin{subfigure}[b]{\columnwidth}
    \centering
    \includegraphics[width=0.8\columnwidth]{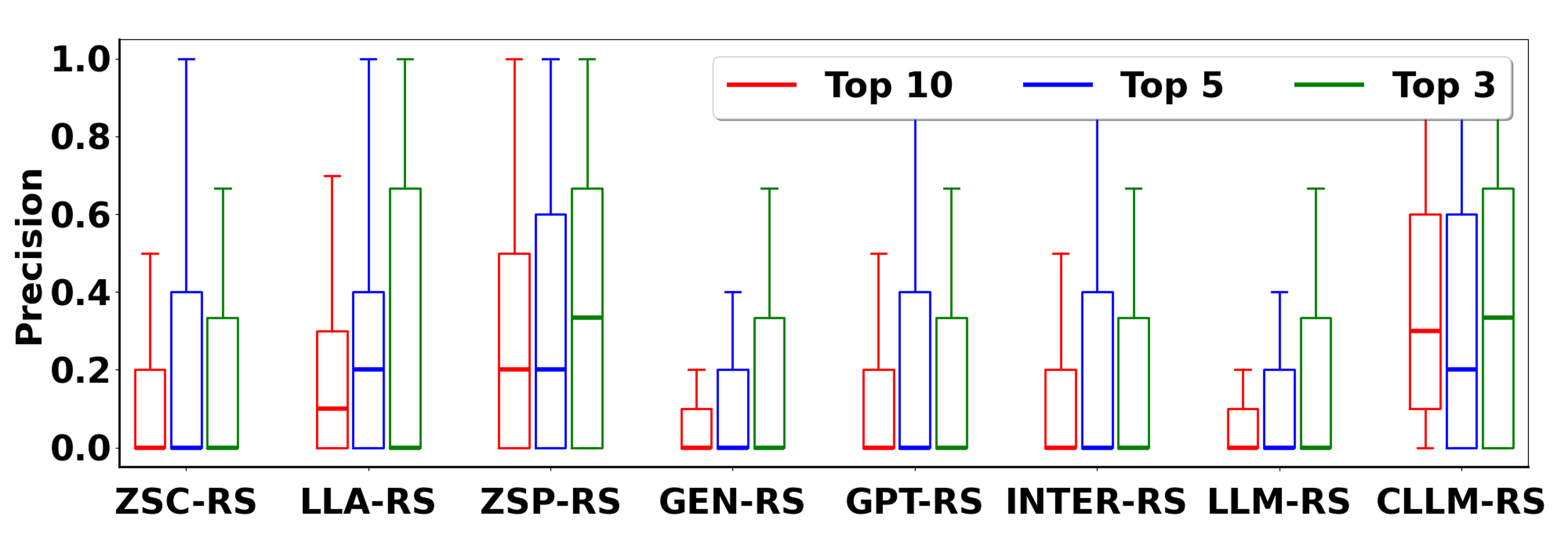}
\end{subfigure}

\begin{subfigure}[b]{\columnwidth}
    \centering
    \includegraphics[width=0.8\columnwidth]{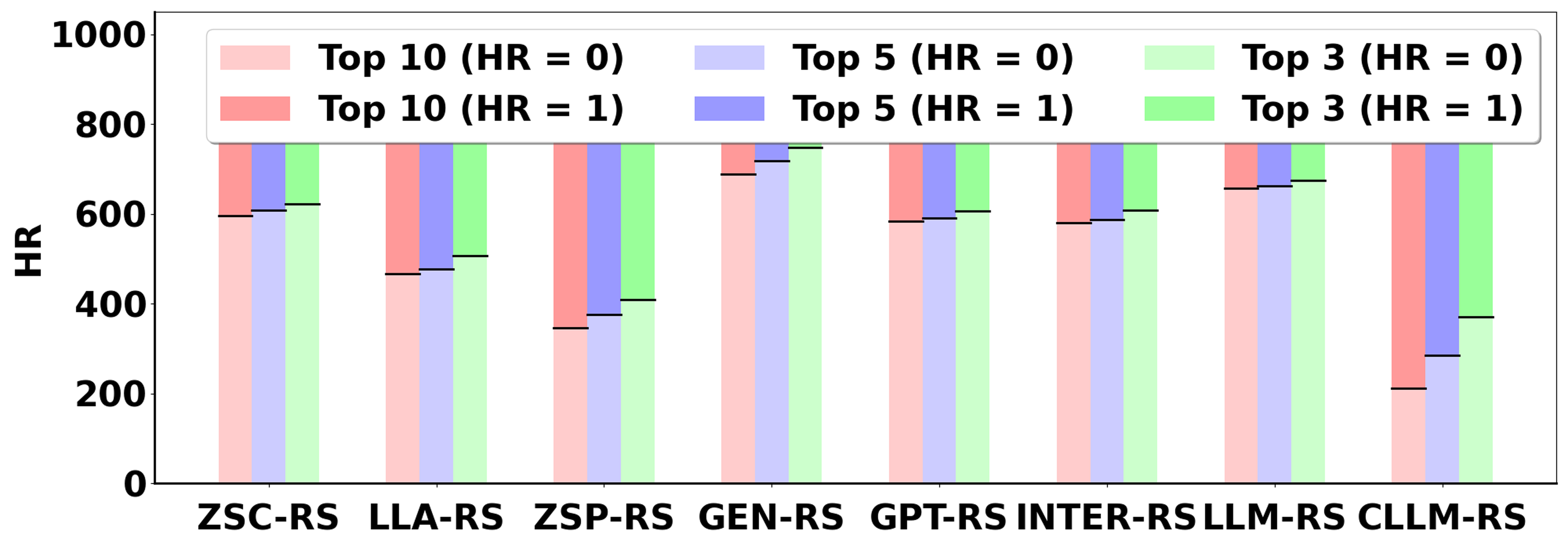}
\end{subfigure}

\begin{subfigure}[b]{\columnwidth}
    \centering
    \includegraphics[width=0.8\columnwidth]{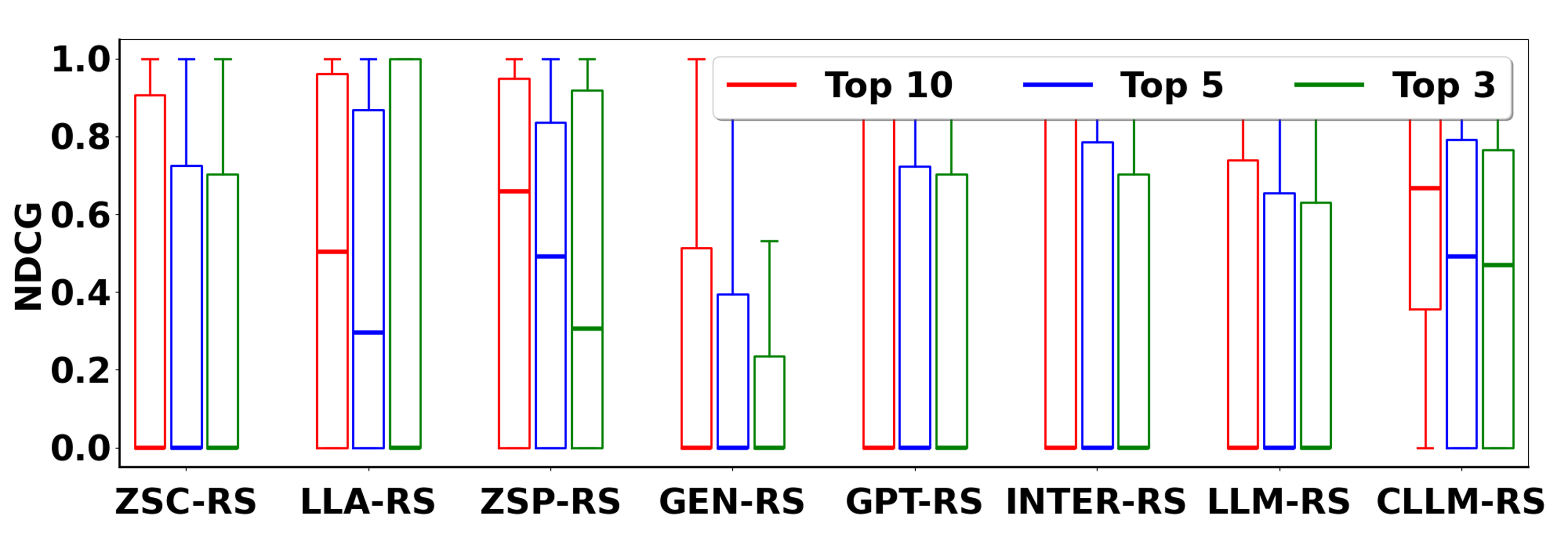}
\end{subfigure}
\caption{Evaluation on Movies with a candidate set.}
\label{fig:movie_critic_candidate}
\end{figure}



\section{Experiment}\label{Experiment}
We evaluate the performance of Critic-LLM-RS against the state-of-the-art LLM-as-RS. Experiments are conducted on a cluster where each node has 8 cores, 64G RAM, and an NVIDIA GeForce RTX 3090. We employ a popular open-source large language model called Vicuna \cite{zheng2024judging} throughout the experiment. Specifically, we use version 1.5 of the model with a size of 7B parameters. The code will be made publicly available upon acceptance. 

\subsection{Datasets}\label{sec:datasets}
The evaluation is based on two datasets, \textit{Movies} and \textit{Books}. From the \textit{MovieLens Tag Genome dataset of 2021}\cite{kotkov2021revisiting}, the Movies dataset incorporates interactions of 247,383 users with 84,661 movies up to 2020. Given a user and a movie, the interaction includes the movie title, director, main actors, and the user rating. Movie ratings range from 0.5 to 5 with the step of 0.5. From the \textit{2022 book data Tag Genome dataset}\cite{kotkov2022tag}, the Book dataset contains interactions between 350,332 users and 9,374 books up to 2017. Given a user and a book, the interaction includes the book title, URL, authors, language, year published, book description, and user rating. Book ratings range from 1 to 5 with the step of 1. 

By default, $10k$ users in each dataset are randomly selected. 
For each user $u_i$, from the items with real ratings from the user, we randomly select 20 items to form the interaction history and randomly select another item $o_j$; they are the input of the Recommendation Critic as shown in Figure \ref{fig:system}. Among the $10k$ users, 80\% is for training, 10\% for validation, and 10\% for testing.


\subsection{Baselines}
The baselines include the state-of-the-art LLM-as-RS. As discussed in related work, these methods rely on the inherent, pre-trained prowess of LLMs to directly craft recommendations via meticulously designed prompts.
\begin{itemize}
    \item Llama4Rec (LLA-RS) \cite{luo2024integrating} 
    employs data augmentation and prompt augmentation strategies to harness the strengths of both approaches. 
    \item InteraRec (INTER-RS) \cite{karra2024interarec} 
    utilizes Multimodal Large Language Models (MLLMs) to analyze user behavior on web pages, including browsing and interaction data. 
    \item GENREC (GEN-RS) \cite{ji2024genrec} 
    generates personalized recommendation lists by understanding users' historical interactions and preferences through specialized prompts. 
    \item GPTrec (GPT-RS) \cite{petrov2023generative} 
    aims to tackle the challenges of sequential recommendation. 
    \item Zero-shot-Prompt LLM-RS (ZSP-RS) \cite{wei2021finetuned} 
    utilizes natural language commands to help understand complex user needs without targeted training. 
    \item Zero-shot-CoT LLM-RS (ZSC-RS) \cite{kojima2022large} enhances understanding of complex user preferences and logical reasoning by introducing the "think step-by-step" prompt in the recommendation process. 
\end{itemize}

\subsection{Evaluation Metrics}
The evaluation metrics 
are HR(hit rate), NDCG(Normalized Discounted Cumulative Gain), and Precision. 
\begin{itemize}
    \item HR@N measures the proportion of users where at least one relevant item appears in the top-N recommendations. Specifically, it is considered a "hit" if the top-N recommendations to a user contain one item of interest to the user. In this study, rating $\geq 4$ to an item signals the item is in the interest of the user for both Movies and Books datasets. HR@N is defined as the average number of hits across all users:
    \begin{equation}
        \text{HR@N} = \frac{\text{Number of users with hits}}{\text{Total number of users}}
    \end{equation}
    The higher HR@N indicates a better performance. 
    \item 
    NDCG@N evaluates the overall quality of the top-N recommendations and the accuracy of their rankings by assigning higher weights to the items at the top of the rank.
    \begin{equation}
        \text{NDCG@N} = \frac{\text{DCG@N}}{\text{IDCG@N}}
    \end{equation}    
    \begin{equation}
        \text{DCG@N} = \sum_{i=1}^{N} \frac{2^{rel_i} - 1}{\log_2(i+1)}
    \end{equation}
    where $rel_i$ is the relevance score of the recommended item in the $i$-th position from top-$1$ in the rank. In our experiments, $rel_i$ is 1 if the user's rating of the item is $\geq 4$, 0 otherwise. 
    IDCG@N is DCG@N when the top-N recommended items are sorted in the best possible way, i.e., the items with the higher relevance scores are closer to the top of the rank. The higher NDCG@N indicates a better performance.
    \item Precision@N measures the percentage of items in the top-N recommendations that are in the users' interest. In our experiments, if the user's rating of an item is $\geq 4$, the item is in the interest of user; not otherwise. 
    \begin{equation}
        \text{Precision@N} = \frac{\text{Number of items of interest}}{N}
    \end{equation}
    The higher Precision@N indicates a better performance.
\end{itemize}

As shown in Table \ref{tbl:case1}, some recommended items do not have real ratings from users because (i) the items are not in the dataset but recommended by LLM or (ii) the items are in the dataset but the users never rate them. Because HR@N, NDCG@N, and Precision@N require the real ratings of the top-N recommended items, two approaches are adopted in our tests to address the issue. First,
a candidate set of items from datasets is provided with the ratings from a user and we prompt LLM to recommend the user items from the set. 

Second, we train a separate recommendation model, called \textit{Oracle}, similar to the Recommendation Critic but trained using much more data. To understand the performance of the Oracle, the Recommendation Critic is trained using training data of different sizes, and the performance is presented in Table \ref{tbl:criticAcc}. The Recommendation Critic is a multiclass classifier. The rating prediction accuracy of the Critic is measured using micro-averaging precision (ACC) and micro-averaging recall (RECALL), defined as:
\begin{equation} \label{eq1}
\begin{split}
    \text{Accuracy}_{\text{micro-average}} & = \frac{\sum_{i=1}^{L}TP_i}{\sum_{i=1}^{L}(TP_i + FP_i)} \\
    \text{Recall}_{\text{micro-average}} & = \frac{\sum_{i=1}^{L}TP_i}{\sum_{i=1}^{L}(TP_i + FN_i)}    
\end{split}
\end{equation}
where $L$ is the number of classes (i.e., rating levels, 10 for Movies and 5 for Books), $TP_i$, $FP_i$, and $FN_i$ are the true positive, false positive, and false negative for class $i$, respectively. From Table \ref{tbl:criticAcc}, we can observe that the rating prediction is improved dramatically when the training data increases from $3k$ to $10k$. But after $10k$, the performance increase remarkably slows down or even stops. The convergence signals that $10k$ is sufficient for training the Recommendation Critic. The Recommendation Critic trained using $30k$ users is adopted as the Oracle.  

\begin{table}[]
\footnotesize
\centering
\caption{Performance of Recommendation Critic trained using data of different sizes (bold values are the best and underlined values are the second best).}
\label{tbl:criticAcc}
\begin{tabular}{@{}cccccccc@{}}
\toprule
\multicolumn{2}{c}{Model}       & 3k     & 5k     & 10k    & 20k    & 30k \\ \midrule
\multirow{2}{*}{Movie} & ACC    & 0.6797 & 0.6844 & 0.7698 & \underline{0.7702} & \textbf{0.7720} \\
                       & Recall & 0.5469 & 0.5796 & 0.6892 & \underline{0.6903}  & \textbf{0.6913} \\
\multirow{2}{*}{Book}  & ACC    & 0.7106 & 0.7432 & 0.7735 & \underline{0.7746} & \textbf{0.7765} \\
                       & Recall & 0.5734 & 0.6589 & 0.6919 &  \underline{0.6931} & \textbf{0.6952} \\ \midrule
\end{tabular}
\end{table}

%
\subsection{Critic-LLM-RS vs Baselines}
The performance of Critic-LLM-RS (denoted as CLLM-RS) and Baselines are reported in Figure \ref{fig:book_baselines} and Figure \ref{fig:movie_baselines} for top-3, top-5, and top-10 recommendations where the evaluations are based on Oracle if real ratings are not available. Besides baselines, we also compare with LLM-RS which represents the initial recommendations of Critic-LLM-RS. The performance of our Critic-LLM-RS is significantly better than all baselines and LLM-RS in most cases on both datasets. The dominant performance of Critic-LLM-RS against LLM-RS verifies the effectiveness of the Recommendation Critic. Without the feedback from the Recommendation Critic, the initial recommendations (i.e., LLM-RS) do not have an advantage compared with baselines. 

When a candidate set of items from datasets is provided with the ratings from a user, we prompt LLM to recommend the user items from the set. In this situation, the performance of Critic-LLM-RS and Baselines are reported in Figure \ref{fig:book_critic_candidate} and Figure \ref{fig:movie_critic_candidate}. A similar performance advantage of our Critic-LLM-RS can be observed compared with baselines and LLM-RS.

\begin{figure}[t]
  \centering
  \includegraphics[width=1\linewidth]{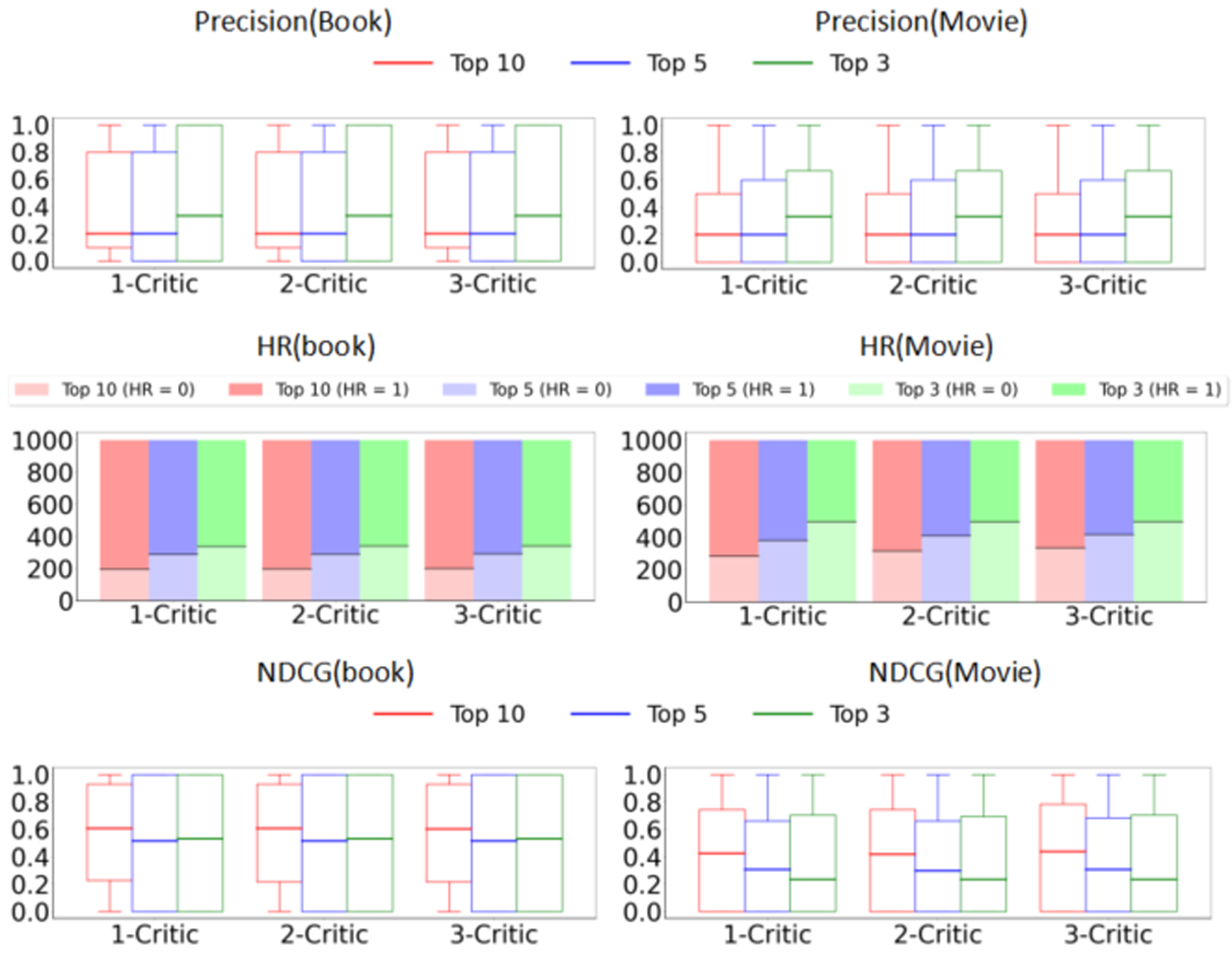}
  \caption{Provide feedback in multiple loops.}
  \label{fig:book_movie_critic}
\end{figure}




\subsection{Feedback in Multiple Loops}
So far, the Recommendation Critic provides feedback on the recommendations of LLM, and then LLM takes the feedback to recommend again. This is called a loop. What if this is repeated for multiple loops? The test results are reported in Figure \ref{fig:book_movie_critic}. Interestingly, the performance is almost the same for one loop, two loops, and three loops. This demonstrates that LLM can effectively learn how to refine the recommendations in a single loop.

\subsection{Critic-LLM-RS vs Fine-tuned Method}
To get a better insight into the performance of Critic-LLM-RS, we managed to fine-tune the LLM\footnote{We use LoRA~\cite{hu2021lora} for the supervised fine-tuning} using item interactions history of users to build the input,  titles of items with a true rating of 4 or higher are used as the expected output. The input of the LLM shares the same prompt template for initial recommendations in Table \ref{tbl:case1}. The LLM used in supervised fine-tuning is Vicuna, for the hyperparameters of supervised fine-tuning, we set the learning rate as 1e-4 and set the training epochs as 3. 

\begin{figure}[t]
  \centering
  \includegraphics[width=1\linewidth]{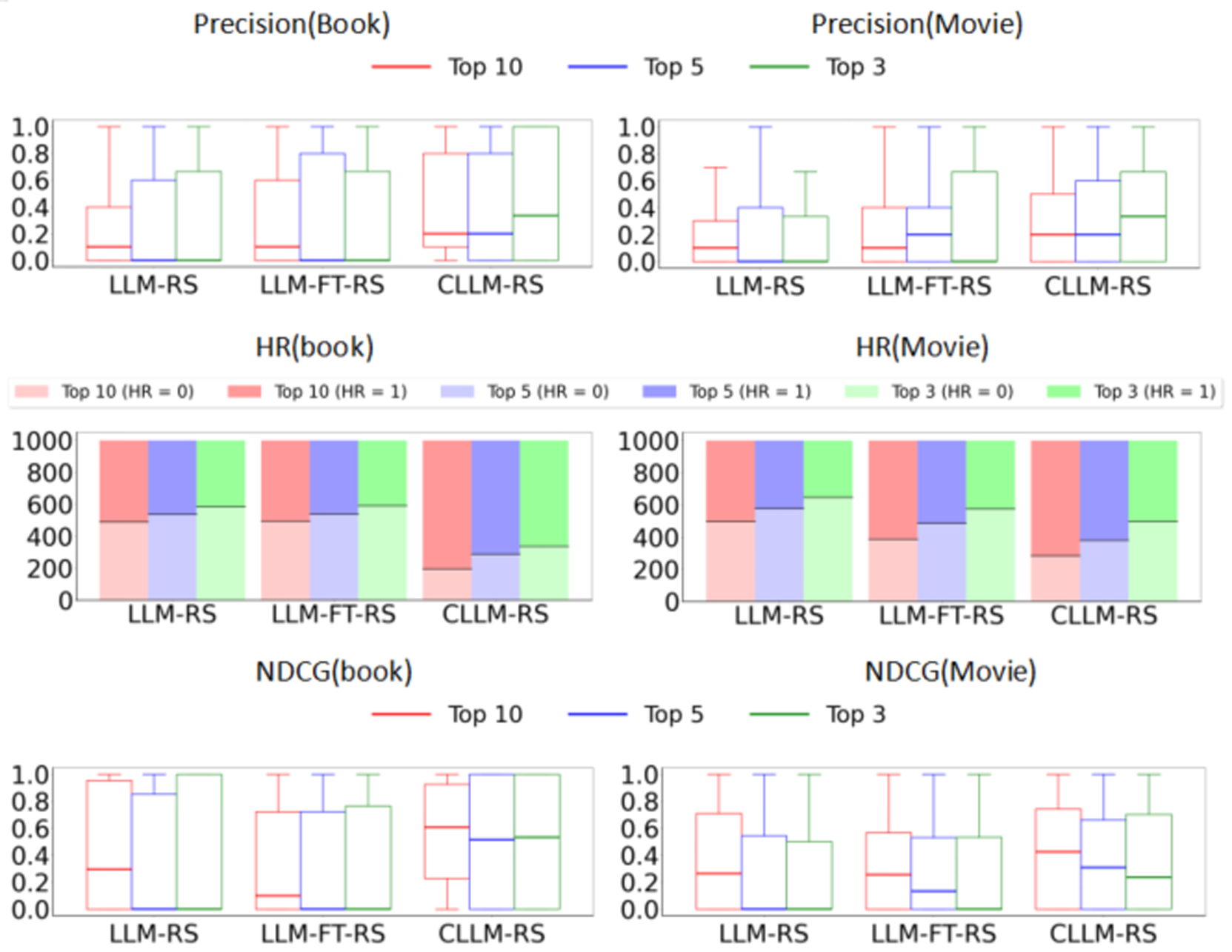}
  \caption{Critic-LLM-RS vs Fine-tuned Method}
  \label{fig:book_movie_ft}
\end{figure}

Figure \ref{fig:book_movie_ft} shows the experiment results of the directly supervised fine-tuning of the LLMs. It is clear that compared to LLM-RS, both LLM-FT, which enhances recommendations through fine-tuning, and Critic-LLM-RS, which optimizes recommendations through iterative feedback, exhibit superior performance. Notably, Critic-LLM-RS performs even better, suggesting that fine-tuning may not be the optimal approach for improving recommendation quality, while feedback-driven optimization offers significant advantages in this regard. Look closely, this might be caused by the loss function when performing supervised fine-tuning in LLMs, the loss function is a token-by-token match targeting to output the ground truth text when provided corresponding input, which deviates from the objective of a recommendation system. 







\subsection{Critic-GPT4o vs GPT4o Method}
\begin{figure}[t]
  \centering
  \includegraphics[width=1\linewidth]{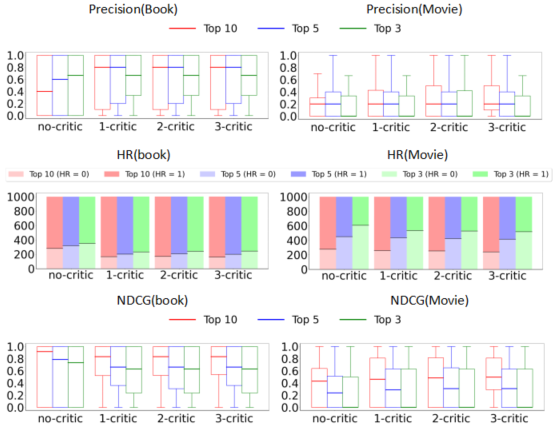}
  \caption{Critic-GPT4o-RS vs GPT4o-RS Method}
  \label{fig:book_movie_gpt4o}
\end{figure}

To comprehensively evaluate the performance of the Critic model in feedback adjustment and optimization of LLM recommendations, we applied it to the large black-box LLM model GPT4o to further enhance the quality of its generated recommendations. The input to GPT4o follows the same prompt template as in Table \ref{tbl:case1} for initial recommendations. Figure \ref{fig:book_movie_gpt4o} illustrates the changes in recommendation performance before and after applying Critic to GPT4o, as well as the results after one, two, and three rounds of feedback adjustment. 

On Figure \ref{fig:book_movie_gpt4o}, no-critic denotes the GPT4o-RS without Critic for feedback to adjust the recommendation results, while 1-critic, 2-critic, and 3-critic denote the Critic-GPT4o-RS model with Critic added for feedback of the recommendation results for one round of feedback, two rounds of feedback, and three rounds of feedback. The figure clearly shows that the recommendation performance of GPT4o improves significantly after incorporating Critic for feedback adjustment. This demonstrates that the feedback optimization approach of the Critic model is equally effective for large black-box LLM models, providing an innovative and effective method for further enhancing the recommendation performance of complex models.
\subsection{Critic's Impact on LLM Invocation Time}
\begin{figure}[t]
\centering
\begin{subfigure}[t]{0.45\columnwidth} 
    \centering
    \includegraphics[width=\textwidth]{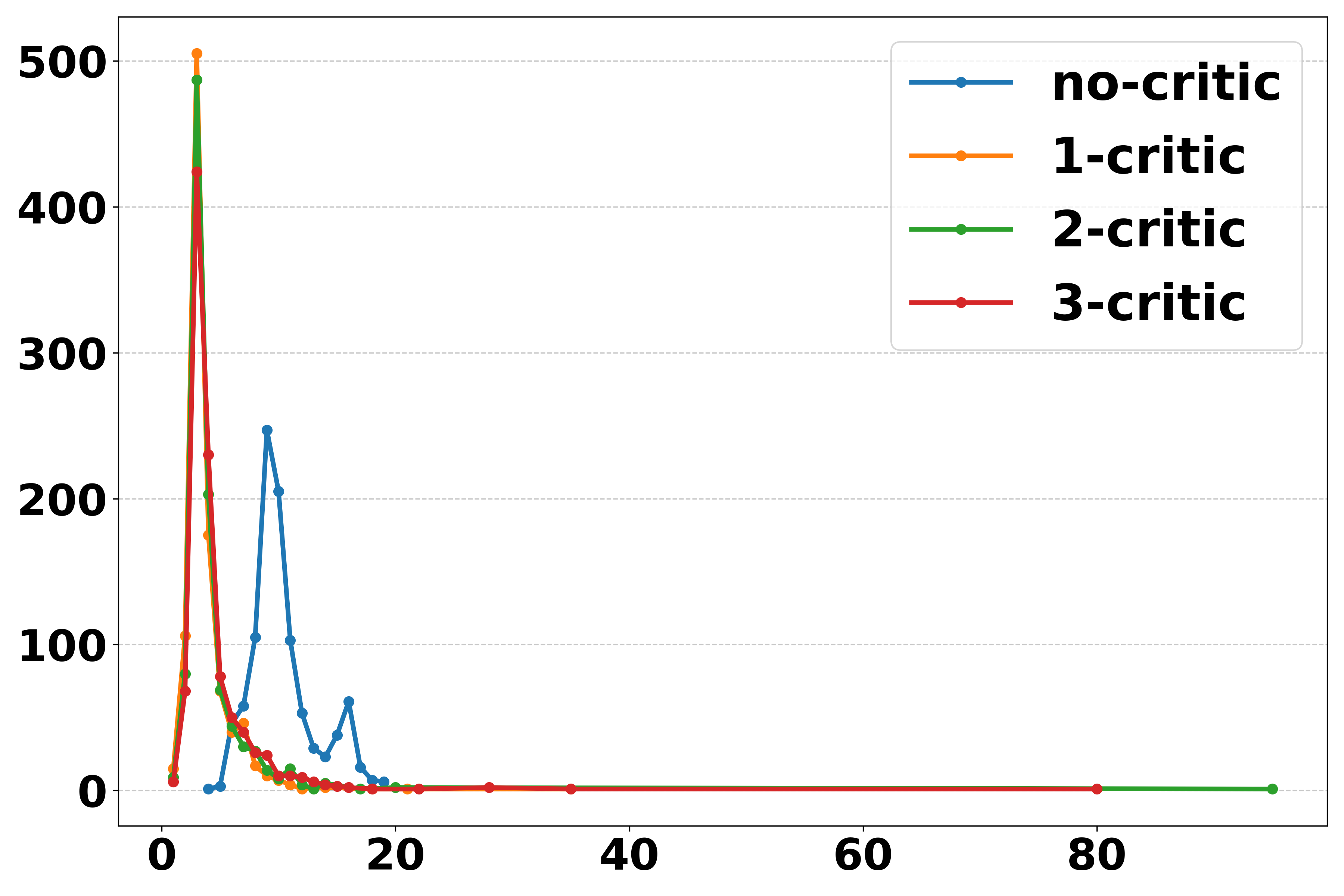}
    \caption{Book} 
    \label{fig:gptvbook}
\end{subfigure}
\hfill 
\begin{subfigure}[t]{0.45\columnwidth} 
    \centering
    \includegraphics[width=\textwidth]{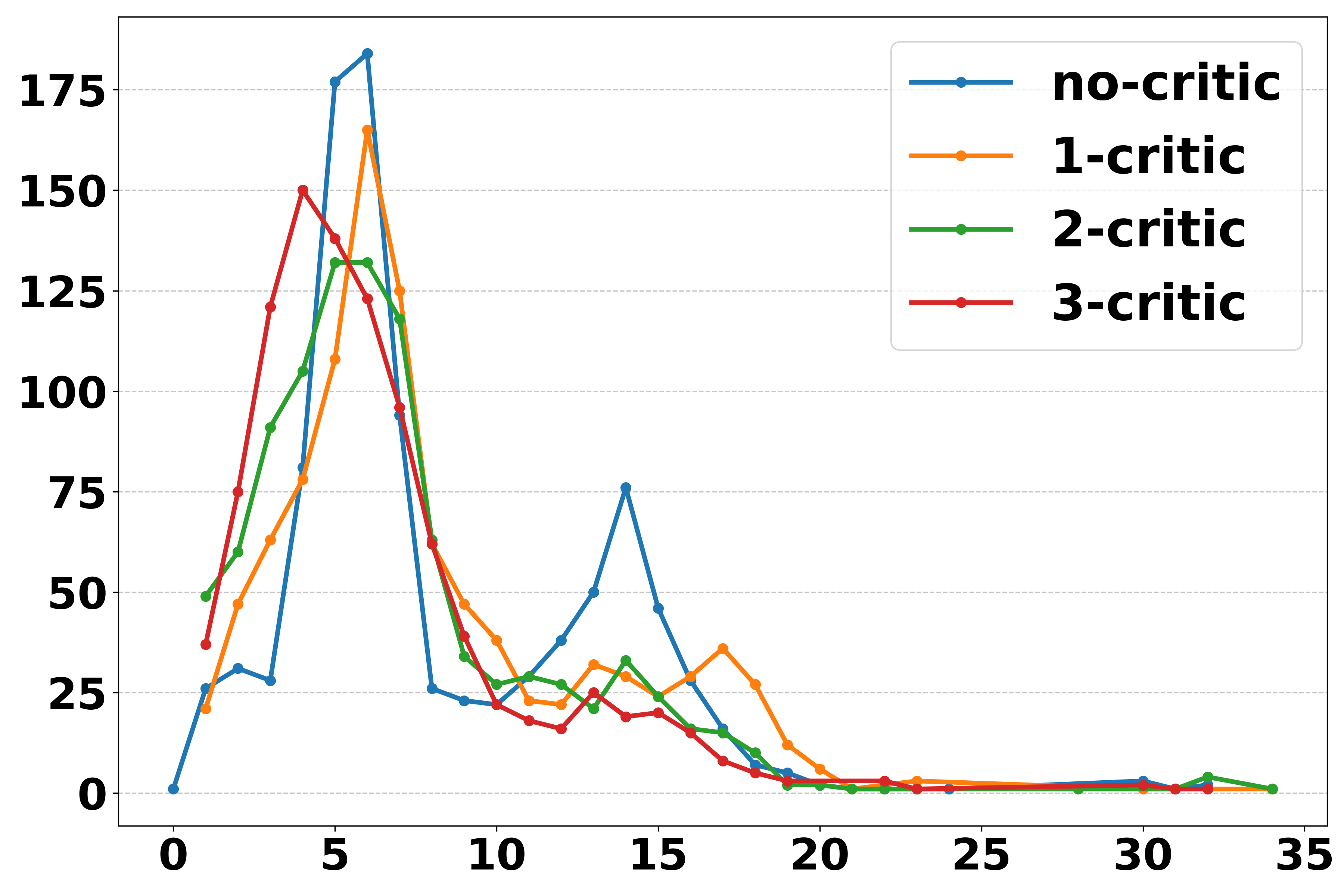}
    \caption{Movie} 
    \label{fig:gptvmovie}
\end{subfigure}
\caption{Critic-LLM-RS vs LLM-RS}
\label{fig:gptv}
\end{figure}

\begin{figure}[t]
\centering
\begin{subfigure}[t]{0.45\columnwidth} 
    \centering
    \includegraphics[width=\textwidth]{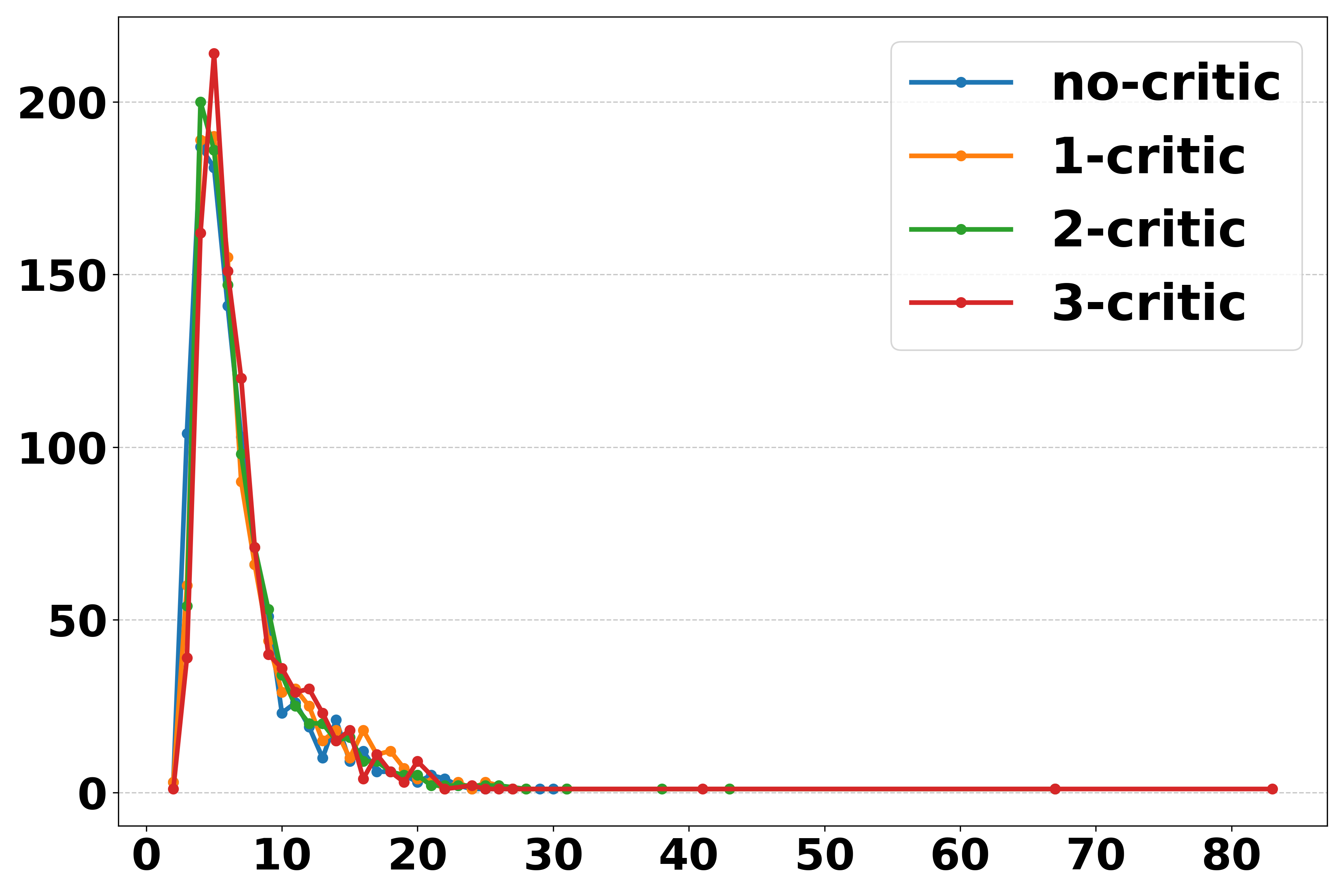}
    \caption{Book} 
    \label{fig:gpt4obook}
\end{subfigure}
\hfill 
\begin{subfigure}[t]{0.45\columnwidth} 
    \centering
    \includegraphics[width=\textwidth]{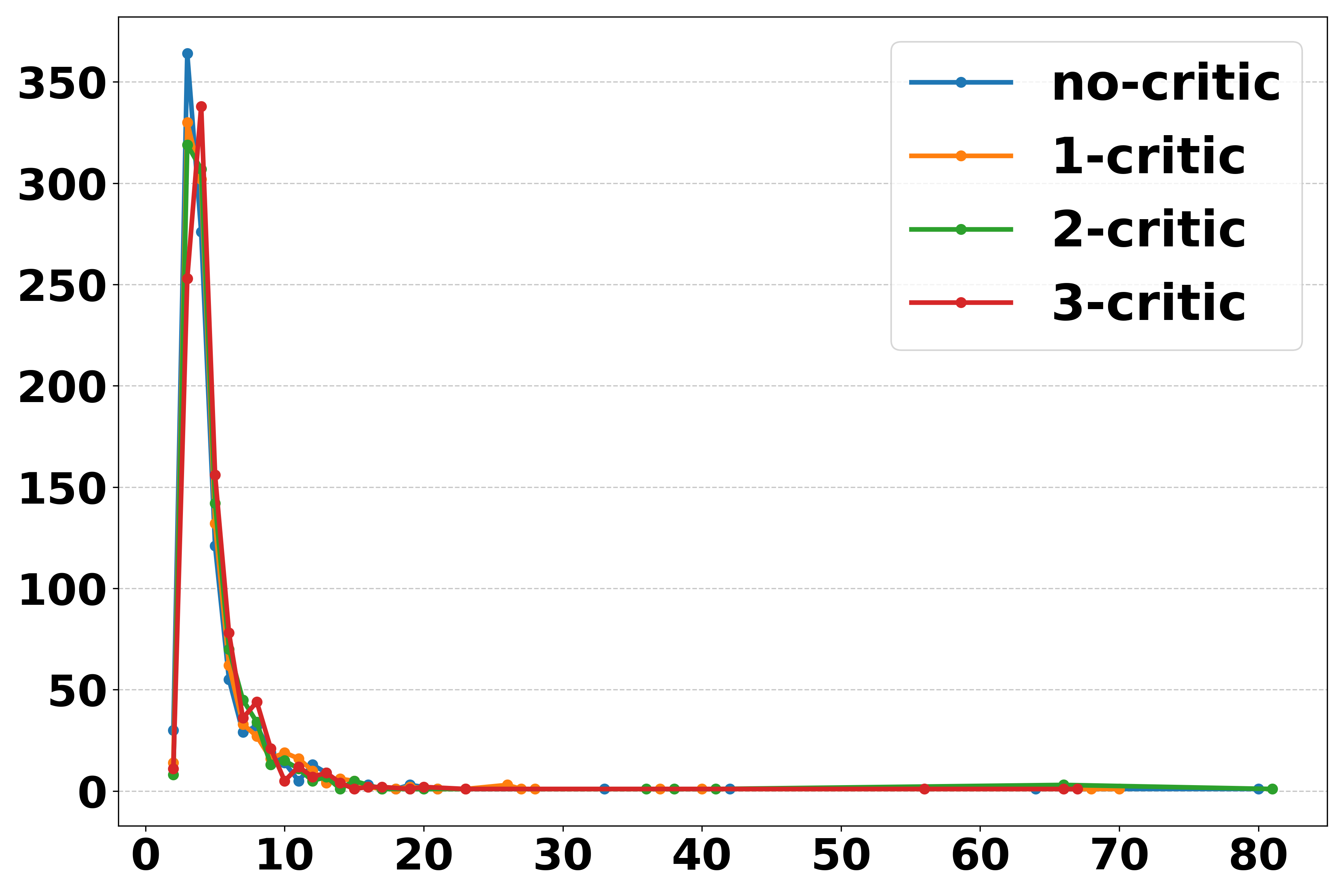}
    \caption{Movie} 
    \label{fig:gpt4omovie}
\end{subfigure}
\caption{Critic-GPT4o-RS vs GPT4o-RS}
\label{fig:gpt4o}
\end{figure}

To provide a more intuitive demonstration of Critic's performance and verify its impact on LLM invocation time, we analyzed the time distribution of LLM calls on Vicuna and GPT4o before and after adding Critic, as well as after one, two, and three rounds of feedback with Critic. Figures \ref{fig:gptv} and \ref{fig:gpt4o} illustrate the time distribution of LLM calls before and after using Critic on Vicuna and GPT4o, as well as the time distribution for different rounds of feedback after adding Critic.

The results show that, whether on Vicuna or GPT4o, the distribution of LLM invocation times exhibits no significant changes before and after using Critic. This indicates that while Critic effectively enhances recommendation performance, it introduces minimal additional burden in terms of LLM invocation time and resource consumption. This finding further validates Critic's ability to improve recommendation performance while maintaining high efficiency.

\section{Conclusion}
This study proposes to train a separate machine-learning model called \textit{Recommendation Critic} for the capability of collaborative filtering and integrates it with LLM as a recommendation system, named Critic-LLM-RS. 
By providing LLM feedback on its recommendations, Critic-LLM-RS prompts LLMs to refine recommendations. 
Critic-LLM-RS enjoys the comprehensive knowledge of pre-trained LLM which can recommend items not included in the recommendation system training dataset, but also enjoys the accurate recommendations due to collaborative filtering. The extensive experiments and a case study have verified the effectiveness of the Critic-LLM-RS for recommendation tasks.

\bibliography{aaai25}

\end{document}